\documentclass[12pt]{article}
\usepackage{latexsym}
\begin{document}
\newcommand{\roughly}[1]%
       
\newcommand\lsim{\roughly{<}}
\newcommand\gsim{\roughly{>}}
\newcommand\CL{{\cal L}}
\newcommand\CO{{\cal O}}
\newcommand\half{\frac{1}{2}}
\newcommand\beq{\begin{eqnarray}}
\newcommand\eeq{\end{eqnarray}}
\newcommand\eqn[1]{\label{eq:#1}}
\newcommand\intg{\int\,\sqrt{-g}\,}
\newcommand\eq[1]{eq. (\ref{eq:#1})}
\newcommand\meN[1]{\langle N \vert #1 \vert N \rangle}
\newcommand\meNi[1]{\langle N_i \vert #1 \vert N_i \rangle}
\newcommand\mep[1]{\langle p \vert #1 \vert p \rangle}
\newcommand\men[1]{\langle n \vert #1 \vert n \rangle}
\newcommand\mea[1]{\langle A \vert #1 \vert A \rangle}
\def\Dsl{\,\raise.15ex \hbox{/}\mkern-12.8mu D}
\newcommand\Tr{{\rm Tr\,}}
\thispagestyle{empty}
\begin{titlepage}
\begin{flushright}
{CALT-68-2307} \\
\end{flushright}
\vspace{1.0cm}
\begin{center}
{\LARGE \bf  Universal Asymptotic Behavior of}\\ 
\bigskip
{\LARGE \bf Mortgage Prepayments }\\
~\\
\bigskip\bigskip
{ Mark B. Wise$^a$ and Vineer Bhansali$^b$} \\
~\\
\noindent
{\it\ignorespaces
           (a) California Institute of Technology, Pasadena CA 91125\\
           
          {\tt wise@theory.caltech.edu}\\
\bigskip   (b) PIMCO, 840 Newport Center Drive, Suite 300\\
               Newport Beach, CA 92660 \\
{\tt   bhansali@pimco.com}
}\bigskip
\end{center}
\vspace{2cm}
\begin{abstract}
 Mortgage prepayments play a crucial role in the pricing and hedging of mortgage backed
securities. An important feature of mortgage prepayment modeling is burnout; as time goes on those
borrowers who have the greatest tendency to refinance are removed from the pool leaving only those that
are less likely to refinance. In this paper we examine the implications of burnout on the late time
prepayment rate using rather general assumptions. Analytic formulas are derived for the average prepayment rate
in the $N$'th month, $P_N^{SMM}$, and the fraction of borrowers remaining in the pool in the $N$'th month,
$y_N$. In the case where the incentive to refinance, and other relevant economic factors, are constant
these results are particularly simple. For example, $P_N^{SMM}=p^{(0)}+(1-p^{(0)})/N+...$, where $p^{(0)}$ is a
constant and the ellipses denote terms suppressed by more powers of $1/N$ or exponentially suppressed.
The term of order $1/N$ indicates that burnout causes the probability of prepayment to decrease as a very simple function of $N$.

\end{abstract}
\vfill
\today
\end{titlepage}

The pricing and price volatility of mortgage backed securities depend critically on the mortgage prepayment rate (Fabozzi [1992]). Prepayment
models have several components, of which the two most important are refinancing and turnover. The refinancing component of a prepayment model captures prepayments due to the incentive provided by lower mortgage rates in the market than the mortgage holder's own mortgage. The turnover component captures prepayments due to mortgagor mobility, home sales etc. An
important aspect of the refinancing component is that mortgage holders in the pool have different
propensity to refinance. As time goes on those with the greatest tendency to refinance are removed from
the pool leaving only those less likely to refinance. This phenomena is usually called {\it burnout} (Hayre [1994]). In this
paper we examine the implications of burnout for the prepayment rate at late times. Assuming
that the refinancing probability and the initial distribution of borrowers are smooth functions
of the borrowers propensity to refinance, we derive formulas for the average single monthly mortality in the $N$'th 
month, $P_N^{SMM}$, and the fraction of borrowers left in the pool after the $N$'th month, $y_N$ ({\it i.e.} the survival factor). In the case that the
incentive to prepay, and other relevant economic factors, are constant with time these results are
very simple;  $P_N^{SMM}=p^{(0)}+(1-p^{(0)})/N$, and $y_N=y_{N_0}(N_0/N)(1-p^{(0)})^{(N-N_0)}$, where $p^{(0)}$ is a
constant. (Here $N$, $N_0$ and $N-N_0$ are assumed large and terms less important for
large values of these quantities are neglected.) The term of order $1/N$ in $P_N^{SMM}$ reflects the fact that
burnout causes the prepayment rate to decrease with $N$ and it gives rise to the factor of $N_0/N$ in the expression
for $y_N$.
We compare our prediction for $P_N^{SMM}$ with the results of simulations based on the PIMCO prepayment
model and find good agreement when $N$ is large.

Suppose the pool of mortgage holders is distributed in a parameter $\theta$ that characterizes their
propensity to refinance. The (normalized)
distribution of mortgage holders in the pool at the end of month $n$ is denoted by
 $f_n(\theta)$, and the initial
distribution of mortgage holders in the pool is denoted by $f_0(\theta)$. The integral
of $f_n(\theta)$ over allowed values of $\theta$ is unity. Let $p_r(n;\theta)$ be the probability of refinancing,
at a given $\theta$, in month $n$. It depends on various factors of which the most
important is the incentive to refinance. The incentive to refinance is variously modeled by different groups to be a function of the difference, or the ratio of the mortgage coupon and the current mortgage rate. The probability of prepaying in month $n$
due to other effects like turnover and defaults is denoted by $p_o(n)$. At a given value of $\theta$, the total probability of prepaying in the 
$n$'th month is $p(n;\theta)=p_r(n;\theta)+p_o(n)$.

Without loss of generality we assume that $f_0(\theta)=1$ with $0\leq \theta\leq 1$. If this isn't true just
choose a new parameter $\theta'$ to label the propensity to refinance such that $d\theta '=f(\theta )d\theta$. In terms
of this variable the probability of prepaying in month $n$ is $p(n; \theta')=p(n;\theta(\theta'))$. The choice $f_0(\theta)=1$ has the disadvantage that it makes the probability to prepay dependent on the initial distribution of borrowers
in propensity to refinance, but it is convenient for our purposes. Then the distribution of borrowers after month $n$ is (Hayre, Chaudhary and Young [2000]), 
\beq 
f_n (\theta )= A_n(1 - p (n;\theta )) \ldots (1 - p (1;\theta ))~, 
\eqn{evolve}
\eeq
where the normalization constant $A_n$ is determined from the condition that the integral of $f_n(\theta)$ over $\theta$ is unity. By assuming that $f_n$ satisfies \eq{evolve}
we have put the media effect into the value of $p(n;\theta)$ (rather than in the evolution of $f_n(\theta)$) and neglected changes in $f_n(\theta)$ arising from the evolution of borrower credit and financial circumstances. 

Suppose larger values of $\theta$ correspond to a greater propensity to
refinance. Then if the incentive to refinance is significant $p(n;\theta)$ is an increasing function of $\theta$ and 
\eq{evolve} implies that the distribution of borrowers becomes peaked near $\theta =0$ for late times. Since at late
times it is the small $\theta$ region that is relevant we expand the probability of prepaying in a power series
about $\theta=0$,
\beq 
p(n;\theta)=p^{(0)}(n)+p^{(1)}(n)\theta + \ldots~,
\eqn{pser}
\eeq
where the ellipses denote terms with higher powers of $\theta$. Clearly \eq{pser} relies on the assumption that
 the
 probability $p(n;\theta)$ is smooth in $\theta$ so that it can be expanded in a power series. The first
term in \eq{pser} is  the prepayment probability at $\theta=0$,  
$p^{(0)}(n)=p_r(n;0)+p_o(n)$, and the coefficient of $\theta$ in the 
second term is the first derivative of the refinancing probability with
respect to $\theta$ evaluated at $\theta=0$, $p^{(1)}{(n)}=dp_r(n;0)/d\theta$. Combining \eq{evolve}
and the terms explictly shown in \eq{pser} and using
\beq
(1-p(n;\theta))=(1-p^{(0)}(n))\exp\left[-{p^{(1)} (n)\theta \over 1 - p^{(0)}(n)}+\ldots\right]~,
\eqn{method}
\eeq
gives that, for large $N$, the distribution of borrowers after month $N$ is,
 \beq
f_N(\theta)=N \delta_N \exp[-N \delta_N \theta]~,
\eqn{func}
\eeq 
where
\beq
\delta_N={1 \over N}\sum_{n=1}^N \left[{p^{(1)} (n)\over 1 - p^{(0)}(n)}\right]~.
\eqn{delt}
\eeq
Here we have assumed that $p^{(1)}(n) \neq 0$. If $p^{(1)}(n)$ is zero or anomalously small then terms we have
considered as subdominant for large $N$ are actually important.

Since $\delta_N$ is the average of $N$ terms it is reasonable to assume (provided the refinancing incentive
remains significant for most of the time) that $\delta_N$ is order unity ({ $\it i.e.$ is not
small for large $N$). Then \eq{func} indicates
that as time evolves from the origination of the mortgage to month $N$ the distribution of mortgage holders, in propensity to refinance, goes from flat (by convention) to
strongly exponentially peaked near $\theta=0$. Note that \eq{func} is valid only for large $N$ and
 terms suppressed by powers of $1/N$ or $\exp[-N]$ have been neglected. It is not likely that 
for early months the first two terms in the power series expansion of the prepayment rate about $\theta =0$ are the dominant ones. However, for large $N$, these are a negligible part of the sum in \eq{delt}.

The average prepayment rate in the $n$'th month is (Hayre, Chaudhary and Young [2000])
\beq
P^{SMM}_n=\int_{0}^{1}p(n;\theta)f_{n-1}(\theta)d\theta~.
\eqn{smm}
\eeq
Using the expression for $f_N$ in \eq{func}, and \eq{pser}, the average single month mortality 
at late times is given by,
\beq
P_N^{SMM} =p^{(0)} (N) + {p^{(1)} (N)\over N \delta_{N-1}} +\ldots~,
\eqn{smn}
\eeq
where the ellipses represent terms suppressed by more powers of $1/N$ or exponentially suppressed.
Thus burnout causes the prepayment rate to evolve towards $p^{(0)}(N)$, differing from it, by the term of order $1/N$
shown in \eq{smn}. In the case where $p^{(0)}(n)=p^{(0)}$ and $p^{(1)}(n)=p^{(1)}$ are constants independent of time
the average prepayment rate in the $N$'th month becomes,
\beq
P_N^{SMM} =p^{(0)} + {1-p^{(0)}\over N}  +\ldots~.
\eqn{beauty}
\eeq
All dependence on the first derivative of the prepayment probability with respect to $\theta$, $p^{(1)}$, has
disappeared. The prepayment rate $P_N^{SMM}$ approaches $p^{(0)}$ in a universal fashion independent
of the details of the initial distribution of borrowers and the $\theta$ dependence of
the probability to prepay. 

Prepayment rates are of often quoted on a yearly basis. Using the relation 
$1-P_N^{CPR}=(1-P_N^{SMM})^{12}$, between
the constant prepayment rate $P^{CPR}_N$ and the single monthly mortality, \eq{beauty} implies that
\beq
P_N^{CPR} =1-(1-p^{(0)})^{12} + {12(1-p^{(0)})^{12}\over N}  +\ldots~.
\eqn{beauty1}
\eeq
Eqs. (8) and (9) are the main results of this paper.

It is interesting to compare  \eq{beauty} with what would be obtained using the continuous time approximation. Then \eq{evolve} becomes, 
\beq
f_N(\theta)=A_N \exp \left[-\int_{0}^{N} p(n;\theta)~dn\right].
\eqn{conttime}
\eeq
This gives a coefficient of the order $1/N$ term in \eq{beauty}
of $1$ instead of $1-p^{(0)}$. The $ p^{(0)}$ in the order $1/N$ term is arises from the
fact that the payments are made discretely and not continuously.

The validity of \eq{beauty} or \eq{beauty1} does not require the prepayment probability to be independent of
time for all time. Only the late time behavior is important in the derivation of these results. These simple formulas may provide a useful``rule of thumb" for estimating the importance of burnout on the late time prepayment rate. 

The value of N at which \eq{beauty1} becomes an accurate approximation can be ascertained by
comparing this equation with the results of a realistic prepayment model. In Figure 1 we compare this
equation with the results of a simulation based on the PIMCO prepayment model 
for generic Fannie Mae 30-year mortgages with low seasoning.
The effects of turnover were shut off and the incentive
to refinance was a constant 200 bp, to emphasize long time burnout behavior. The darker line
is \eq{beauty1} with the value $p^{(0)}=0.00958$ fitted to the large $N$ part of the simulation. 
We see that \eq{beauty1} fits very well after
$N=100$. Fits of similar quality hold for other values of the incentive to refinance and if turnover is included.

Our results depend crucially on the assumption that the mortgage holders are smoothly distributed in the
continuous variable $\theta$. Suppose
that $\theta$ could only take on a few discrete values, $\theta_k$, $k=1,2, \ldots,k_{max}$,
and the propensity to refinance increases with $k$. Let the initial fraction of borrowers corresponding
to $\theta_k$ be $f_k$ and let $p(\theta_k)$ denote their prepayment
probability. In this case the analog of \eq{beauty} for the late time behavior of
the prepayment rate is (Hayre [1994])}
\beq
P_N^{SMM} =p(\theta_1) + (p(\theta_2)-p(\theta_1)){f_2 \over f_1}\left[{1-p(\theta_2) \over 1-p(\theta_1)}\right]^{N-1}  +\ldots~.
\eqn{ugly}
\eeq
Now burnout causes the prepayment rate to approach $p(\theta_1)$ exponentially fast. Furthermore,
the approach is not universal depending on $f_2/f_1$ and $p(\theta_2)-p(\theta_1)$. 

With a large pool of mortgage
holders it is likely that treating $\theta$ as a continuous variable is appropriate. 
Even with $\theta$ continuous it is
possible that the refinance probability is not analytic in $\theta$. This would occur, for example, if a finite
fraction of the pool had zero probability to prepay even if the incentive to prepay was significant. Empirically, mortgage
holders that never prepay are rare and it seems reasonable that in a statistical sense they can
be treated as a set of measure zero. 

The fraction of mortgage holders remaining in the pool after the n'th month is 
\beq
y_n =(1-P_n^{SMM}) \ldots (1-P_1^{SMM})~.
\eqn{yevolve}
\eeq
Using \eq{beauty} for the single month mortality \eq{yevolve} gives,
\beq
y_N = y_{N_0}(N_0/N)(1-p^{(0)})^{(N-N_0)}~,
\eqn{ybeauty}
\eeq
where it is assumed that, $N$, $N_0$ and $N-N_0$ are all large. The factor of $N_0/N$ is a consequence of burnout.
Since the single month mortality and the fraction of borrowers left in the pool both depend on the month $N$, we
can eliminate $N$ and express the single month mortality as a function of the fraction remaining 
in the pool, $P^{SMM}(y)$. For small $y$ \eq{beauty} and \eq{ybeauty} imply that,
\beq
P^{SMM}(y)=p^{(0)}+{(1-p^{(0)}) \ln (1-p^{(0)})\over \ln y} +...~,
\eqn{pfy}
\eeq
where the ellipses denote terms less important as $y$ goes to zero. Even though the refinancing probability 
was assumed to be an analytic function of the propensity to refinance the average  prepayment rate is
not analytic as a function of $y$. It's derivatives are singular at $y=0$. We do not expect the
term explicitly displayed in \eq{pfy}
to be an accurate approximation unless $y$ is very small, since neglecting 
the ellipses treats $-\ln y$ as much larger than
$\ln (-\ln y)$. Also, keeping only the term explicitly displayed in
\eq{pfy} is clearly not valid in the case $p^{(0)}=0$. In that case, $P^{SMM}(y)= \xi y$, where $\xi= 1/N_0 y_0$ is a constant.

In this paper we examined the implications of burnout for the late time behavior of the 
single monthly mortality $P_N^{SMM}$ and the survival factor $y_N$ (in the $N$'th month). Our main
assumption was that the probability to
refinance and the initial distribution of borrowers are smoothly distributed in a variable, $\theta$,
 that labels borrowers
propensity to refinance. In the case that the
incentive to prepay, and other relevant economic factors, are constant with time these results are
very simple;  $P_N^{SMM}=p^{(0)}+(1-p^{(0)})/N$, and $y_N=y_{N_0}(N_0/N)(1-p^{(0)})^{(N-N_0)}$, where $p^{(0)}$ is a
constant. (Here $N$, $N_0$ and $N-N_0$ are assumed large and terms less important for
large values of these quantities are neglected.) The term of order $1/N$ in $P_N^{SMM}$ reflects the fact that
burnout causes the prepayment rate to decrease with $N$ and it gives rise to the factor of $N_0/N$ in the expression for $y_N$. The approach of the single monthly mortality to its limiting value, $p^{(0)}$, is independent
of the details of the initial distribution of borrowers and the $\theta$ dependence of the probability of
refinancing. We compared our prediction for $P_N^{SMM}$ with results from simulations
based on the  PIMCO prepayment
model and found reasonable agreement when $N$ is large.

The authors would like to thank Lakhbir Hayre at Salomon Smith Barney for his contributions to the prepayment modeling approach relevant to this paper. They would also like to thank Bill Campbell and the portfolio analytics team at PIMCO for their help in running numerical simulations. 

\vspace{0.4cm}

\noindent
{\Large{\bf References}}

\vspace{0.4cm}

\noindent
Frank J. Fabozzi, {\it The Handbook of Mortgage Backed Securities}, Probus Publishing Company, Chicago IL, 1992.

\vspace{0.4cm}

\noindent
Lakbir S. Hayre, {\it A Simple Statistical Framework for Modeling Burnout and Refinancing Behavior}, The Journal of
Fixed Income, Volume 4, Number 3, December 1994.

\vspace{0.4cm}

\noindent
Lakhbir S. Hayre, Sharad Chaudhary and Robert Young, {\it Anatomy of Prepayments: The Saloman Smith Barney Prepayment Model}, Salomon Smith Barney, April 2000.

\vspace{0.4cm}
\noindent
 {\Large{\bf Figure Caption}}
\vspace{0.4cm}


\noindent
Figure 1. Plot of CPR in percent versus month. The grey line is the result of a simulation
 based on the PIMCO prepayment model with a constant 200 bp refinancing incentive and no turnover. The black
line is \eq{beauty1} with the fitted value $p^{(0)}=0.00958$.





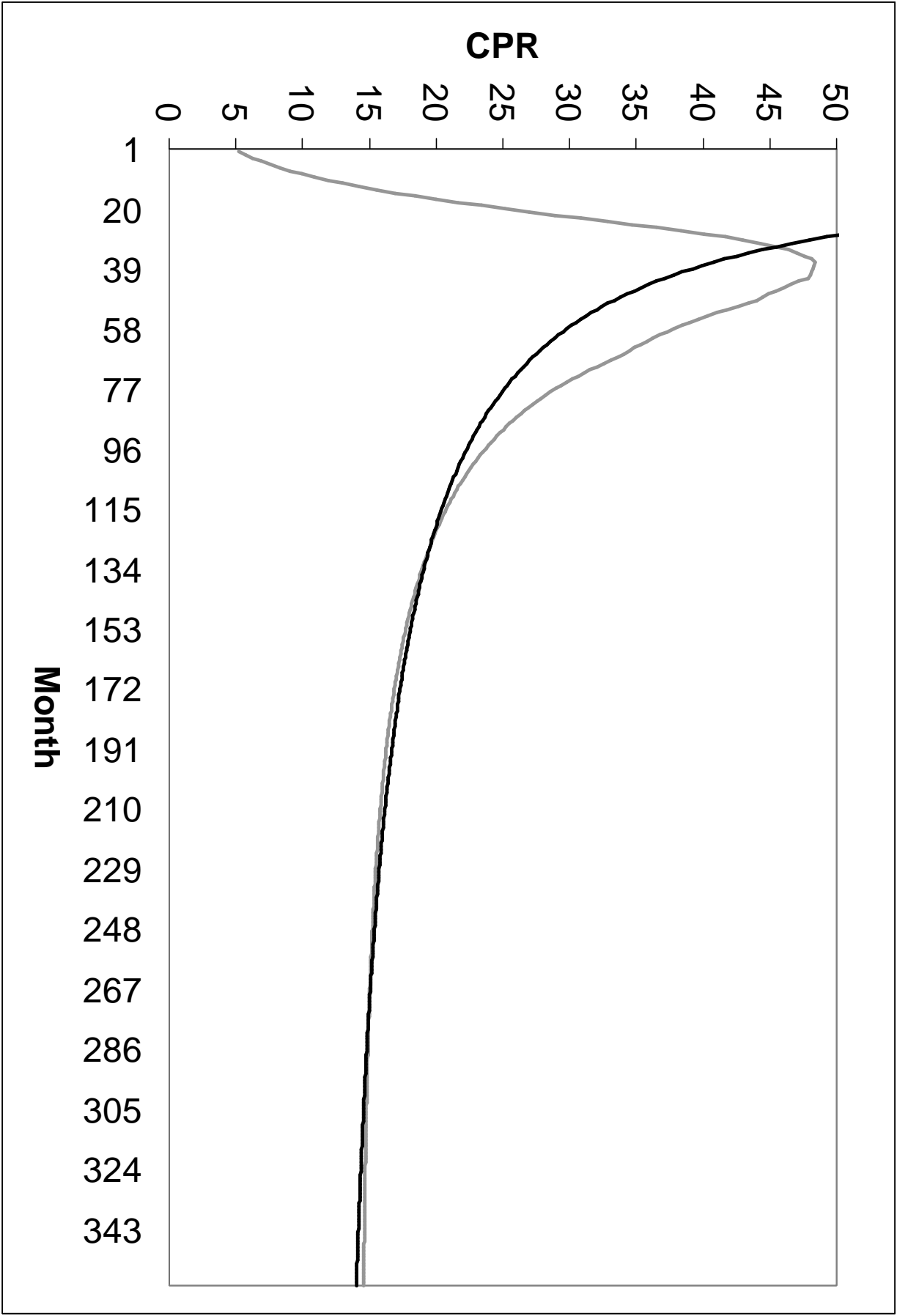
\end{document}